\documentclass[preprint]{ptephy}
\preprintnumber{YITP-16-101, KUNS-2639}

\usepackage{boites}
\usepackage{graphicx}
\usepackage{dcolumn}
\usepackage{bm}
\usepackage{braket}
\usepackage{setspace} %
\usepackage{comment} 
\usepackage{color}

\usepackage{version}	
\begin{document}

\title{ 3$\alpha$-cluster structure and monopole transition in $^{12}$C and $^{14}$C }

\author{\name{\fname{Yuta} \surname{Yoshida}}{1\ast} and \name{\fname{Yoshiko} \surname{Kanada-En'yo}}{2} }


\address{
\affil{1}{Yukawa Institute for Theoretical Physics, Kyoto University, Kyoto 606-8502, Japan}
\affil{2}{Department of Physics, Kyoto University, Kyoto 606-8502, Japan}
\email{yyuta@yukawa.kyoto-u.ac.jp}
}

\begin{abstract}
3$\alpha$-cluster structures and monopole transitions of $0^+$ states in $^{12}$C and $^{14}$C were investigated with $3\alpha$- and $^{10}$Be+$\alpha$-cluster models.
A gas-like $3\alpha$ state and a bending-chain $3\alpha$ state were obtained in the $0^+_2$ and $0^+_3$ states of $^{12}$C, respectively. 
In $^{14}$C, a linear-chain 3$\alpha$ structure is found in the $0^+_4$ state near the $^{10}$Be+$\alpha$ threshold, but a cluster gas-like state does not appear because valence neutrons attract $\alpha$ clusters and suppress spatial development of 3$\alpha$ clustering.
It was found that the linear-chain state in $^{14}$C is stabilized against the bending and $\alpha$ escaping modes by valence neutrons. 
The monopole transition strengths in $^{12}$C are enhanced by $3\alpha$-cluster developing, whereas, those in $^{14}$C are not enhanced so much because of the tight binding of $\alpha$ clusters by valence neutrons.
\end{abstract}

\subjectindex{}

\maketitle



\section{INTRODUCTION}
\label{sec:INTRO}

Cluster structure is one of the important aspects in nuclear systems, in particular in light nuclei including unstable nuclei.
One of the typical examples of cluster structures is a $3\alpha$ cluster structure in $^{12}$C.
The 3$\alpha$ cluster in $^{12}$C has been intensively investigated theoretically and experimentally \cite{Uegaki-12C,Fujiwara1980,Kamimura-12C,Descouvemont1987,THSR,Fedotov2004,Filikhin2005,Kurokawa-12C,Suhara-12C,Ohtsubo2013,Freer2014,Ishikawa2014,Funaki-12C,Enyo-12C}.
In the 1950s, a linear-chain $3\alpha$ state was suggested by Morinaga to describe the $0^+_2$ state in $^{12}$C near the $3\alpha$ threshold energy \cite{Morinaga1956,Morinaga1966}.
However, this idea was excluded at least for the $0^+_2$ state by the experimental $\alpha$-decay width larger than the expectation for the linear chain state.
Later, the $0^+_2$ state was understood as a gas-like $3\alpha$ state \cite{Uegaki-12C,THSR,Funaki-12C}.
The large monopole transition strength between the ground and $0^+_2$ states supports the gas-like $3\alpha$ structure in the $0^+_2$.
Generally, monopole transitions are receiving a lot of attention because they are good probes to identify cluster structures in excited states \cite{Yamada:2011ri, Funaki:2006gt,Kawabata:2005ta,KanadaEn'yo:2006bd,Wakasa:2006nt,Ichikawa:2011ji,Chiba:2015zxa}.
For $^{12}$C, various cluster structures other than the gas-like state have been also suggested in an energy region higher than the $0^+_2$ state.
For instance, the bending-chain $3\alpha$ state was predicted by Antisymmetrized Molecular Dynamics (AMD) and Fermionic molecular dynamics (FMD) calculations \cite{Neff2004,Enyo2007,Suhara2010}.

For neutron-rich C isotopes, possibilities of linear-chain $3\alpha$ structures have been discussed in many theoretical and experimental works \cite{Itagaki2001,Soic2003,vonOertzen2004,Price2007,Haigh2008,Maruhn2010,Suhara-14C,Baba2014,Freer2014a,Fritsch2016}.
One of the major interests is whether excess neutrons stabilize linear-chain $3\alpha$ structures in neutron-rich C.
For $^{14}$C, the $\beta$-$\gamma$ constraint AMD calculation has predicted a linear-chain $3\alpha$ structure and its rotational band at an energy slightly higher than the $^{10}$Be+$\alpha$ threshold energy \cite{Suhara-14C}.
The predicted linear-chain $3\alpha$ structure shows a $^{10}$Be+$\alpha$ cluster structure, in which valence neutrons attract 2$\alpha$ clusters form the $^{10}$Be core and the third $\alpha$ locates at the head-on position of the deformed $^{10}$Be core.
In recent experiments, the $2^+$ and $4^+$ states have been reported by $^{10}$Be scattering on $\alpha$ and regarded as candidates for members of the predicted linear-chain band in $^{14}$C \cite{Fritsch2016}.
Also for $^{16}$C, a linear chain structure was predicted by an AMD calculation \cite{Baba2014}.

Our aim in this paper is to theoretically investigate cluster structures of $0^+$ states in $^{12}$C and $^{14}$C, and discuss their contributions to monopole transitions.
For this aim, we adopt 3$\alpha$- and $^{10}$Be+$\alpha$-cluster models.
We apply the generator coordinate method (GCM) with these models, and investigate $3\alpha$ dynamics without and with valence neutrons in $^{12}$C and $^{14}$C.
In this work, we particularly focus on $\alpha$-cluster motion around Be cores and $\alpha$-$\alpha$ motion in the Be cores to discuss appearance and stabilization of gas-like and linear-chain $3\alpha$ structures.
Roles of valence neutrons in $^{14}$C are also discussed. 

This paper is organized as follows.
In Sec. \ref{sec:framework}, we explain the present framework of the $3\alpha$- and the $^{10}$Be+$\alpha$-cluster models with the GCM.
The calculated results of $^{12}$C and $^{14}$C are shown in Sec. \ref{sec:results}.
In Sec. \ref{sec:discussion}, we analyze details of cluster structures and discuss their contributions to monopole transition strengths.
Finally, in Sec. \ref{sec:Summary}, a summary is given.


\section{FRAMEWORK}
\label{sec:framework}

\subsection{3$\alpha$-cluster model}

To describe cluster structures of $^{12}$C, we perform the 3$\alpha$-cluster GCM calculation using the Brink-Bloch $\alpha$-cluster wave function \cite{Brink}.
The Brink-Bloch $\alpha$-cluster wave function for $3\alpha$ is written as follows, 
\begin{eqnarray}
{\rm \Phi}_{3\alpha}({\bf R}_1, {\bf R}_2, {\bf R}_3)
   &=& 
{\mathcal A} [\Phi_{\alpha}({\bf R}_1) \Phi_{\alpha}({\bf R}_2) \Phi_{\alpha}({\bf R}_3)], \\
\Phi_{\alpha}({\bf R}_k;1,2,3,4) 
   &=&
       \varphi_{p \uparrow}({\bf R}_k;1)\varphi_{p \downarrow}({\bf R}_k;2)
       \varphi_{n \uparrow}({\bf R}_k;3)\varphi_{n \downarrow}({\bf R}_k;4), \\
\varphi_{\sigma}({\bf R}_k;i)
   &=& \left(\frac{2\nu}{\pi}\right)^{\frac{3}{4}}
       \exp \left[ -\nu\left({\bf r}_i-{\bf R}_k\right)^2 \right]
       \chi_{\sigma}(i)
       \tau_{\sigma}(i),
\end{eqnarray}
where $\mathcal{A}$ is the antisymmetrizing operator for all nucleons.
$\Phi_{\alpha}({\bf R}_k)$ is the $\alpha$-cluster wave function expressed by the harmonic oscillator $(0s)^4$ configuration with a width parameter $\nu$, which is localized around the position ${\bf R}_k$.
$\chi_\sigma$ and $\tau_\sigma$ are the spin and isospin parts of the single-particle wave function.

\begin{figure}[htbp]
\begin{center}
\includegraphics[width=10cm]{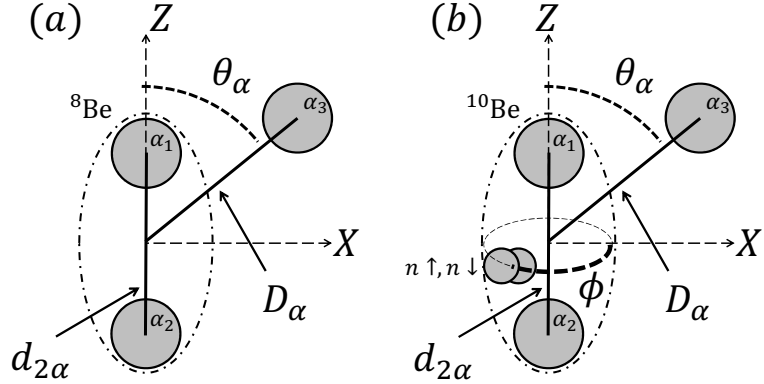}
\end{center}
\caption{\label{fig:model}
Schematic figures for the 3$\alpha$- and $^{10}{\rm Be}+\alpha$-cluster models.
$d_{2\alpha}$ indicates the $\alpha$-$\alpha$ distance, and $D_{\alpha}$ and $\theta_{\alpha}$ indicate the $\alpha_3$ position relative to the center of mass position of the $2\alpha$.
In the right panel for the $^{10}{\rm Be}+\alpha$-cluster model, $\phi$ indicates the angle of the $nn$ position on the $X$-$Y$ plane.
}
\end{figure}

In this work, we focus on the motion of the third $\alpha$ cluster ($\alpha_3$) around the $2\alpha$, therefore, we rewrite the generator coordinates ${\bf R}_1, {\bf R}_2, {\bf R}_3$ with the $\alpha$-$\alpha$ distance ($d_{2\alpha}$) and the distance ($D_{\alpha}$) and angle ($\theta_{\alpha}$) of the $\alpha_3$ position (${\bf D}_{\alpha}$) relative to the center of mass position of the $2\alpha$ as 
\begin{eqnarray}
{\bf D}_\alpha&\equiv& {\bf R}_3-\frac{{\bf R}_1+{\bf R}_2}{2}=(D_\alpha \cos \theta_\alpha, 0, D_\alpha \sin \theta_\alpha)\\ 
{\bf R}_1&=&-\frac{1}{3}{\bf D}_\alpha+\frac{d_{2\alpha}}{2}{\bf e}_z\\
{\bf R}_2&=&-\frac{1}{3}{\bf D}_\alpha-\frac{d_{2\alpha}}{2}{\bf e}_z\\
{\bf R}_3&=&+\frac{2}{3}{\bf D}_\alpha,
\end{eqnarray}
as shown in Fig. \ref{fig:model} (a).
Here, we denote $x$- $y$- and $z$- components of the relative position ${\bf D}_\alpha$ as $(D_{\alpha x},D_{\alpha y},D_{\alpha z})=(X,Y,Z)$.
We rewrite the $3\alpha$ wave function ${\rm \Phi}_{3\alpha} (d_{2\alpha},D_{\alpha},\theta_{\alpha})$ and consider these parameters $\{d_{2\alpha},D_{\alpha}$, $\theta_{\alpha}\}$ as the generator coordinates instead of $\{{\bf R}_1, {\bf R}_2, {\bf R}_3\}$.

To calculate the energy spectra and monopole transitions in $^{12}$C, we superpose the parity and total angular-momentum projected wave functions as 
\begin{eqnarray}
\Psi_{^{12}{\rm C}({0^+_n})}
  &=& \sum_{d_{2\alpha}} \sum_{D_{\alpha},\theta_{\alpha}} c_{0^+_n}({d_{2\alpha},D_{\alpha},\theta_{\alpha}}) \hat{P}^{0+}_{00}
\Phi_{3\alpha}(d_{2\alpha},D_{\alpha},\theta_{\alpha}),
\end{eqnarray}
where $\hat{P}^{J\pm}_{MK}$ is the parity and total-angular-momentum projection operator.
The coefficients $c_{0^+_n}$ are determined by diagonalizing Hamiltonian and norm matrices.

\subsection{$^{10}{\rm Be}+{\alpha}$-cluster model}

In order to describe cluster structures in $^{14}$C, we perform a GCM calculation of $3\alpha$-$nn$ by extending the $3\alpha$-cluster GCM.
In this paper, we pay a particular attention to $^{10}{\rm Be}+\alpha$-cluster structures in $^{14}$C, and therefore, the $nn$ configuration is optimized to describe the $^{10}$Be cluster in $^{14}$C as follows. 
We start from the Brink-Bloch $3\alpha$-cluster wave function ${\rm \Phi}_{3\alpha}({\bf R}_1, {\bf R}_2, {\bf R}_3)$ described previously.
We express two valence neutrons with Gaussian wave packets as $\varphi_{\uparrow n}({\bf R}_{n\uparrow})\varphi_{\downarrow n}({\bf R}_{n\downarrow})$.
Here, the Gaussian centers ${\bf R}_{n\uparrow}$ and ${\bf R}_{n\downarrow}$ for spin-up and down neutrons are chosen so that the subsystem of $2\alpha+nn$ describes a $^{10}$Be cluster as, 
\begin{eqnarray}
{\rm \Phi}_{^{10}{\rm Be}}({\bf R}_1, {\bf R}_2, \phi)
   &=& 
{\mathcal A} \left[ \Phi_{\alpha}\left( {\bf R}_1 \right)\Phi_{\alpha} \left({\bf R}_2\right)
\varphi_{n \uparrow}({\bf R}_{n\uparrow})\varphi_{n \downarrow}({\bf R}_{n\downarrow})
 \right], \\
{\bf Y}_{n\uparrow}&\equiv&{\bf R}_{n\uparrow}
-\frac{{\bf R}_1+{\bf R}_2}{2}=d\left( \cos{\phi}, \sin{\phi},0\right)+i\lambda\left(-\sin{\phi},\cos{\phi} ,0 \right),\\
{\bf Y}_{n\downarrow}&\equiv&{\bf R}_{n\downarrow}
-\frac{{\bf R}_1+{\bf R}_2}{2}=d\left( \cos{\phi}, \sin{\phi},0\right)-i\lambda\left(-\sin{\phi},\cos{\phi} ,0 \right).
\end{eqnarray}
where two neutrons have the Gaussian centers with the same real part (position) and the opposite imaginary part (momentum).
This wave function for the $^{10}$Be cluster is in principle equivalent to the simplified cluster model proposed by Itagaki {\it et al.}\cite{Itagaki}, in which the parameter $\lambda$ for the finite momentum in opposite directions for spin-up and down neutrons is introduced to gain the spin-orbit interaction.
$d$ indicates the distance of $nn$ from the $2\alpha$ and $\lambda$ is the parameter for the momentum of Gaussian wave packets. 
For each $\alpha$-$\alpha$ distance $d_{2\alpha}=| {\bf R}_1- {\bf R}_2|$, we optimize the parameters $d$ and $\lambda$, so as to minimize the subsystem energy $^{10}$Be$(0^+)$ and use the optimized values in the calculation of $^{14}$C.
$\phi$ specifies the angle of the $nn$ position around the $2\alpha$, namely, that from the $3\alpha$ plane in the $^{14}$C wave function (Fig.~\ref{fig:model} (b)).

By using the $^{10}$Be wave function ${\rm \Phi}_{^{10}{\rm Be}}({\bf R}_1, {\bf R}_2, \phi)$, we express the basis wave function for $^{14}$C as
\begin{eqnarray}
 \Phi_{^{10}{\rm Be}+\alpha}({\bf R}_1, {\bf R}_2,{\bf R}_3, \phi)
&=&
{\mathcal A} \left[
{\rm \Phi}_{^{10}{\rm Be}}({\bf R}_1, {\bf R}_2, \phi)
\Phi_{\alpha} \left({\bf R}_3\right) \right],
\end{eqnarray}
where the recoil effect is taken into account as $4{\bf R}_1+4{\bf R}_2+4{\bf R}_3+{\bf R}_{n\uparrow}+{\bf R}_{n\downarrow}=0$.
In a similar way to the $3\alpha$-cluster wave function, we rewrite the generator coordinates ${\bf R}_1, {\bf R}_2, {\bf R}_3$ with $d_{2\alpha}$ for the $\alpha$-$\alpha$ distance and $D_{\alpha}$ and $\theta_{\alpha}$ for the $\alpha_3$ position and express the total wave function as  $\Phi_{^{10}{\rm Be}+\alpha}(d_{2\alpha},D_{\alpha},\theta_{\alpha},\phi)$.

To calculate the energy spectra and monopole transitions in $^{14}$C, we superpose the parity and total angular-momentum projected wave functions by treating four parameters $d_{2\alpha},D_{\alpha},\theta_{\alpha},\phi$ as the generator coordinates as 
\begin{eqnarray}
\Psi_{^{14}{\rm C}({0^+_n})}
  &=& \sum_{d_{2\alpha}} \sum_{D_{\alpha},\theta_{\alpha}}
 \sum_{\phi}  c_{0^+_n}({d_{2\alpha},D_{\alpha},\theta_{\alpha}},\phi) \hat{P}^{0+}_{00}
\Phi_{^{10}{\rm Be}+\alpha}(d_{2\alpha},D_{\alpha},\theta_{\alpha},\phi).
\end{eqnarray}

\subsection{ Isoscalar monopole transition strengths}
The monopole operator $\mathcal{M}({\rm IS0})$ is defined as 
\begin{eqnarray}
\mathcal{M}({\rm IS0}) = \sum_i ({\bf r}_i-{\bf R}_{\rm c.m.})^2,
\end{eqnarray}
where ${\bf r}_i$ is the coordinate of the $i$th particle and ${\bf R}_{\rm c.m.}$ is the total center of mass coordinate.
The monopole transition strength from the ground state to the excite state $0^+_n$ is calculated by the squared matrix element of $\mathcal{M}({\rm IS0})$ as
\begin{eqnarray}
B({\rm IS0};0^+_1 \rightarrow 0^+_n) = |\braket{0^+_1|\mathcal{M}({\rm IS0})|0^+_n}|^2.
\end{eqnarray}

\section{MODEL PARAMETER AND EFFECTIVE INTERACTIONS}
In the present calculation, we use the same width parameter $\nu=0.235$ fm$^{-2}$ used in Ref.~\cite{Suhara-14C}.
For the GCM calculations, discretized values of the generator coordinates are used.
Namely, we take $d_{2\alpha}= 2, 3, 4, 5, 6, 7$ (fm), $D_{\alpha}= 2, 3, 4, 5, 6, 7$ (fm), and $\theta_{\alpha}= 0, \pi/8, \pi/4, 3\pi/8, \pi/2$ in the $3\alpha$ GCM calculation for $^{12}$C. In the $^{10}{\rm Be}+\alpha$ GCM calculation for $^{14}$C, we take $d_{2\alpha}= 2, 3, 4$ fm, $D_{\alpha}= 2, 3, 4, 5, 6$ fm, $\theta_{\alpha}= 0, \pi/8, \pi/4, 3\pi/8, \pi/2$, and $\phi=\frac{\pi}{8},\frac{3\pi}{8},\frac{5\pi}{8},\cdots,\frac{15\pi}{8}$.
It means that we use only three values of $d_{2\alpha}$ for the $^{10}{\rm Be}+\alpha$ GCM calculation to save numerical costs because $d_{2\alpha}$ dependence of the $2\alpha+nn$ subsystem energy shows a deep minimum around $d_{2\alpha}=3$ fm due to valence neutrons. 
More details are described later.

For effective interactions, we adopt the same interactions used in Ref.~\cite{Suhara-14C} for the study of $^{14}$C with the $\beta$-$\gamma$ constraint AMD ($\beta\gamma$-AMD).
We use the Volkov No.2 \cite{Volkov} with $W=1-M$, $B=H=0.125$, and $M=0.60$ for the central force, the spin-orbit term in the G3RS \cite{G3SR} with $u_1=-u_2=1600$ MeV, and the Coulomb force approximated by seven Gaussians.


\section{RESULTS}
\label{sec:results}

\subsection{Energies of $^{8}{\rm Be}$ and $^{10}{\rm Be}$}

\label{sec:results-Be}

\begin{figure}[h]
\begin{center}
\includegraphics[clip,width=9.0cm]{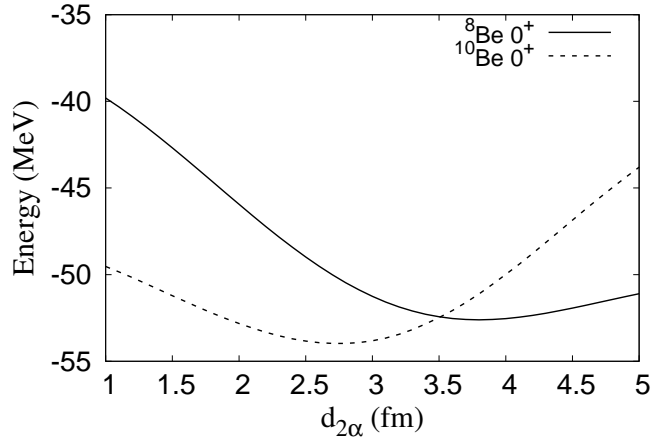}
\end{center}
\caption{\label{fig:Be_EC}
$0^+$-projected energies of $^{8}$Be and $^{10}$Be obtained using the 2$\alpha$ and 2$\alpha+nn$ wave functions, respectively.
The energies are shown as functions of the $\alpha$-$\alpha$ distance $d_{2\alpha}$. 
}
\end{figure}

Here we discuss energies of subsystems, the $^{8}$Be and $^{10}$Be clusters, calculated using the 2$\alpha$- and 2$\alpha$+$nn$-cluster models, respectively.
Figure \ref{fig:Be_EC} shows the $d_{2\alpha}$ dependence of the $0^+$ energy for $^{8}$Be projected from the 2$\alpha$ wave function,
\begin{eqnarray}
{\rm \Phi}_{2\alpha}({\bf R}_1, {\bf R}_2)&=&{\mathcal A} [\Phi_{\alpha}({\bf R}_1) \Phi_{\alpha}({\bf R}_2)],
\end{eqnarray}
and that for $^{10}$Be projected from ${\rm \Phi}_{^{10}{\rm Be}}({\bf R}_1, {\bf R}_2, \phi)$ with $|{\bf R}_1-{\bf R}_2|=d_{2\alpha}$.
In the $0^+$ energy of $^{8}$Be, the energy minimum exists at $d_{2\alpha}=3.8$ fm and the energy curve is found to be very soft toward a large $d_{2\alpha}$ region.
On the other hand, the $0^+$ energy of $^{10}$Be shows the energy minimum at $d_{2\alpha}=2.6$ fm and rapidly increases in a large $d_{2\alpha}$ region. 
It indicates that valence neutrons make $2\alpha$ to be bound tightly and suppress the $\alpha$-$\alpha$ distance in $^{10}$Be compared with $^{8}$Be.

We calculate energy spectra of $^{10}$Be using the GCM within the present 2$\alpha$+$nn$ cluster model by superposing the wave functions ${\rm \Phi}_{^{10}{\rm Be}}({\bf R}_1, {\bf R}_2, \phi)$ with the parameter $d_{2\alpha}=2, 3, 4$ fm and check that the present model reproduces well the spectra of $^{10}$Be($0^+_1$), $^{10}$Be($2^+_1$),  and $^{10}$Be($2^+_2$).

\subsection{Results of $^{12}{\rm C}$}

\subsubsection{Energy surface of 3$\alpha$-cluster structure} \mbox{}\\

\begin{figure}[htbp]
  \begin{center}
   \includegraphics[width=\hsize]{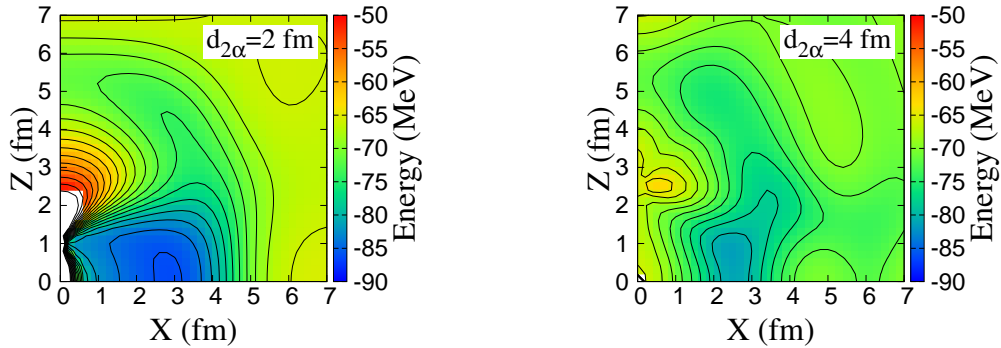}
  \end{center}
 \caption{\label{fig:12C_ES}
$0^+$-projected energy surfaces for the $3\alpha$ wave function $\Phi_{3\alpha}(d_{2\alpha},D_{\alpha},\theta_{\alpha})$ on the $X$-$Z$ plane.
$\alpha$-$\alpha$ distances are chosen to be $d_{2\alpha}=2$ fm and $4$ fm.
}
\end{figure}

Let us discuss the $0^+$-projected energy of the $3\alpha$ wave function $\Phi_{3\alpha}(d_{2\alpha},D_{\alpha},\theta_{\alpha})$. 
Figure \ref{fig:12C_ES} shows the $0^+$-projected energy of the $3\alpha$ wave function plotted on the $X$-$Z$ plane for the $\alpha_3$ position around the $2\alpha$ with $\alpha$-$\alpha$ distances $d_{2\alpha}=2$ fm and $4$ fm.
Here, $X$ and $Z$ are defined as $(X,Z)=(D_\alpha\sin \theta_{\alpha},D_\alpha\cos\theta_{\alpha})$, and the first and second $\alpha$ clusters are located on the $Z$ axis (Fig. \ref{fig:model} (a)).
In the energy surface for $d_{2\alpha}=2$ fm, we find a deep energy pocket around $(X,Z)=(3,0)$ fm which corresponds to the ground state configuration of $^{12}$C.
Along the $Z$ axis, the energy is rather high in the $D_{\alpha}<4$ fm region because of the Pauli blocking effect from another $\alpha$ cluster.
In the $D_{\alpha}\sim 5$ fm region, an energy valley is seen from $\theta_{\alpha}=90^{\circ}$ toward $\theta_{\alpha}=0^{\circ}$ corresponding to the soft bending mode.
In the energy surface for $d_{2\alpha}=4$ fm, the energy surface shows only a shallow minimum around $(X,Z)=(2.5,0)$ fm and an energy plateau is spread widely toward the large $D_\alpha$ region.
It can be seen that the energy does not increase so much with the increase of $d_{2\alpha}$ and $D_\alpha$ indicating that the energy is very soft against the spatial development of the $3\alpha$ cluster except for the energy pocket at the compact $3\alpha$ configuration for the ground state.

\subsubsection{GCM calculation of $^{12}{\rm C}$} \mbox{}\\

\begin{figure}[htbp]
\begin{center}
\includegraphics[width=12.0cm]{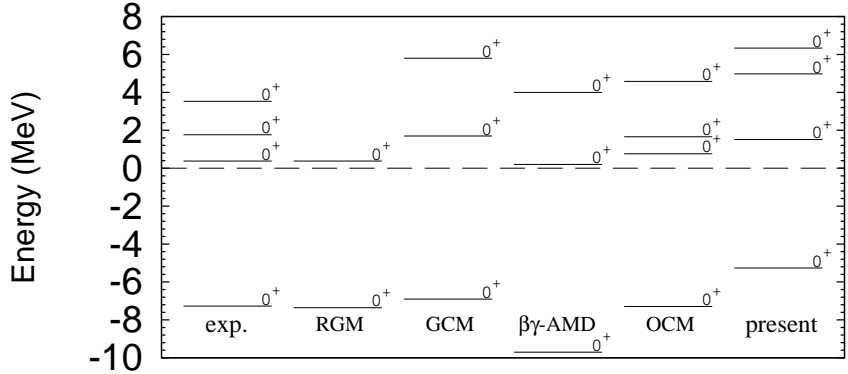}
\end{center}
\caption{\label{fig:level_12C}
$0^+$ energy spectra of $^{12}$C obtained by the Full-GCM calculation of the 3$\alpha$-cluster model measured from the 3$\alpha$ threshold energy.
Experimental spectra \cite{Exp-12C} and theoretical ones of RGM\cite{Kamimura-12C}, GCM\cite{Uegaki-12C}, $\beta\gamma$-AMD\cite{Suhara-12C} and OCM\cite{Kurokawa-12C} are also shown.
}
\end{figure}

We superpose the $3\alpha$ wave functions and obtain the ground and excited $0^+$ states of $^{12}$C with the $3\alpha$ GCM calculation described in Sec. \ref{sec:framework}.
Figure \ref{fig:level_12C} shows the obtained $0^+$ energy spectra compared with the experimental data.
Theoretical spectra calculated using the 3$\alpha$ resonating group method (RGM) by Kamimura {\it et al.} \cite{Kamimura-12C}, the 3$\alpha$ GCM by Uegaki {\it et al.} \cite{Uegaki-12C} the $\beta$-$\gamma$ constraint antisymmetrized molecular dynamics ($\beta\gamma$-AMD) by Suhara {\it et al.}\cite{Suhara-12C}, and the 3$\alpha$ orthogonality condition model (OCM) by Kurokawa {\it et al.}\cite{Kurokawa-12C} are also shown.
These calculations are microscopic $3\alpha$-cluster model calculations except for the OCM and are in principle consistent with the present model but there are slight differences in the interaction parameters and finite volume sizes.
In the present calculation, we obtain the ground state with the compact $3\alpha$ cluster structure, the $0^+_2$ state with the gas-like $3\alpha$ structure, and the $0^+_3$ state with the banding-chain $3\alpha$ structure.
These three states are consistent with those of the $3\alpha$ GCM by Uegaki {\it et al.}

Experimentally, $^{12}$C($0^+_1$, 0 MeV) and $^{12}$C($0^+_2$, 7.65 MeV) are well known.
Around the excitation energy $E_x\sim 10 $ MeV, two $0^+$ states at 9.04 MeV and 10.8 MeV have recently been reported \cite{Itoh-12C}.
However, in the $3\alpha$ GCM and $\beta\gamma$-AMD calculations, only one $0^+$ state is obtained in the $E_x\sim 10 $ MeV region, which can be assigned to the experimental $^{12}$C($0^+$, 10.8 MeV). 
The 3$\alpha$ OCM calculation by Kurokawa {\it et al.} \cite{Kurokawa-12C} and the resent 3$\alpha$ calculation by Funaki {\it et al.} \cite{Funaki-12C} predicted another $0^+$ state of a higher nodal state of the $0^+_2$ state, which may be assigned to the experimental $^{12}$C($0^+$, 9.04 MeV).
In the present calculation, we obtain the $0^+_3$ state which is likely to be assigned to the experimental $^{12}$C($0^+$, 10.8 MeV) state. 
Above the $0^+_3$ state, we obtain the $0^+_4$ state with the higher nodal feature, which might be assigned to $^{12}$C($0^+$, 9.04 MeV).
However, the present model space of $D_\alpha \le 7$ fm is not enough to obtain a converged result for the $0^+_4$ state and could overestimate its excitation energy.

\begin{figure}[htbp]
\begin{center}
\includegraphics[width=\hsize]{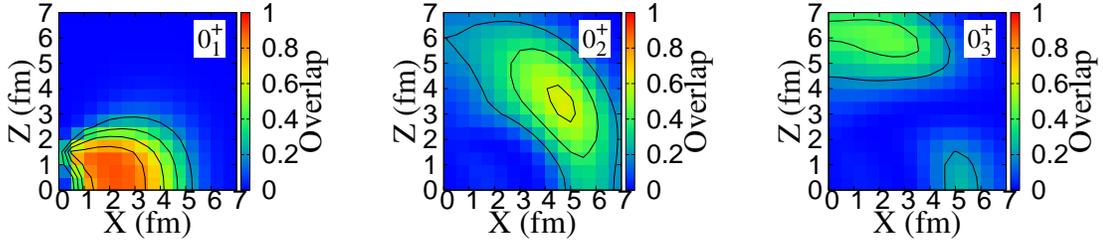}
\end{center}
\caption{\label{fig:12C_Ov}
Overlaps with the $3\alpha$ wave functions ($d_{2\alpha}=3$ fm) for $^{12}$C$(0^+_1)$, $^{12}$C$(0^+_2)$, and $^{12}$C$(0^+_3)$ obtained by the Full-GCM calculation of the $3\alpha$-cluster model.
Overlaps are plotted on the $X$-$Z$ plane for the $\alpha_3$ position.
}
\end{figure}

To discuss $3\alpha$ cluster structures of the obtained $0^+$ states in detail, we calculate the overlap between the GCM wave function for $^{12}$C($0^+_n$) and the basis $3\alpha$ wave function defined as
\begin{eqnarray}
O^{0^+_n}_{^{12}{\rm C}}(d_{2\alpha},D_{\alpha},\theta_{\alpha}) &=& 
\left| \braket{\hat{P}^{0^+}\Phi_{3\alpha} (d_{2\alpha},D_{\alpha},\theta_{\alpha})|
\Psi^{0^+_n}_{^{12}{\rm C}}} \right|^2,
\label{eq:overlap_12C }
\end{eqnarray}
where $|\hat{P}^{0^+}\Phi_{3\alpha}\rangle$ is normalized to be $\langle\hat{P}^{0^+}\Phi_{3\alpha}|\hat{P}^{0^+}\Phi_{3\alpha}\rangle=1$. 
Figure \ref{fig:12C_Ov} shows the overlap plotted on the $X$-$Z$ plane for the $\alpha_3$ position around the $2\alpha$ ($d_{2\alpha}=3$ fm), which indicates the $\alpha_3$ probability distribution in $^{12}{\rm C}(0^+_n)$.
In the ground state, the $\alpha_3$ probability is concentrated around $(X,Z)=(2, 1)$ fm with the maximum value 91 \% showing the compact $3\alpha$-cluster structure. 
In $^{12}$C($0^+_2$), the $\alpha_3$ probability is spread widely in the large $D_{\alpha}$ region in both cases of $d_{2\alpha}=2$ and $4$ fm, indicating the cluster gas-like nature of freely moving 3$\alpha$ clusters like a gas.
This is consistent with the results reported in prior researches \cite{Enyo-12C}.
For $^{12}$C($0^+_3$), the $\alpha_3$ probability distribution shows a large amplitude in $5 {\rm \ fm}<Z<7 {\rm \ fm}$ region near the $Z$-axis corresponding to the aligned $3\alpha$ configuration, but the probability is spread in the finite $X$ region with the maximum amplitude at $(X,Z)=(2.5, 6)$ fm.
Note that the second peak exists around $(X,Z)=(5, 0)$ fm with the amount of 20\%.
These results indicate the bending-chain 3$\alpha$ configuration rather than the linear-chain structure.

\begin{table}[htb]
\caption{  \label{table:12C}
Monopole transition strengths and rms radii of $0^+$ states of $^{12}$C.
Experimental data are from Refs. \cite{Exp-14C,Ozawa2001}.
}
  \centering
  \begin{tabular*}{12cm}{@{\extracolsep{\fill}} c|cccc}
  \hline
   & \multicolumn{2}{c}{Cal.} & \multicolumn{2}{c}{Exp.} \\
\cline{2-3} \cline{4-5}
   State & $B$(IS0) (fm$^4$) & $R_{\rm rms}$ (fm) & $B$(IS0) (fm$^4$) & $R_{\rm rms}$ (fm)\\ 
  \hline
   $0^+_1$  &    & 2.53 &  & 2.35$\pm$0.02 \\ 
   $0^+_2$  & 263 & 3.66 & 120$\pm$10 &  \\ 
   $0^+_3$  & 111 & 3.77 &  &  \\ 
  \hline
  \end{tabular*}
\end{table}

In Table \ref{table:12C}, monopole transition strengths from the ground state and root-mean-square (rms) radii for $^{12}$C$(0^+_n)$ are shown.
The gas-like state ($^{12}$C($0^+_2$)) has the remarkably large monopole transition strength and rms radius.
The $0^+_3$ state of $^{12}$C also has the significant monopole transition strength though it is less than half of that of $^{12}$C($0^+_2$).
The present calculation overestimates the experimental $B$(IS0) by about factor 2 mainly because of the adopted effective interaction.
In the present calculation, we use the original Volkov No.2 interaction as explained previously.
The $3\alpha$ GCM calculation using the Volkov No.1 gives a reasonable $B$(IS0) value \cite{Uegaki-12C}, and the $3\alpha$ RGM using the Volkov No.2 with a modified strength parameter also gives reasonable value for $B$(IS0) \cite{Kamimura-12C}, but the 3$\alpha$ GCM using the original Volkov No.2 overestimates $B$(IS0) \cite{Suhara-12C}.
Although the quantitative difference in $B$(IS0) is found between Volkov No.1 and Volkov No.2 interactions, qualitative behaviors of $3\alpha$-cluster structures are similar between two interactions as shown in the results of Refs.~\cite{Uegaki-12C, Suhara-12C}.

\subsection{Results of $^{14}{\rm C}$}

\begin{figure}[htbp]
  \begin{center}
   \includegraphics[width=\hsize]{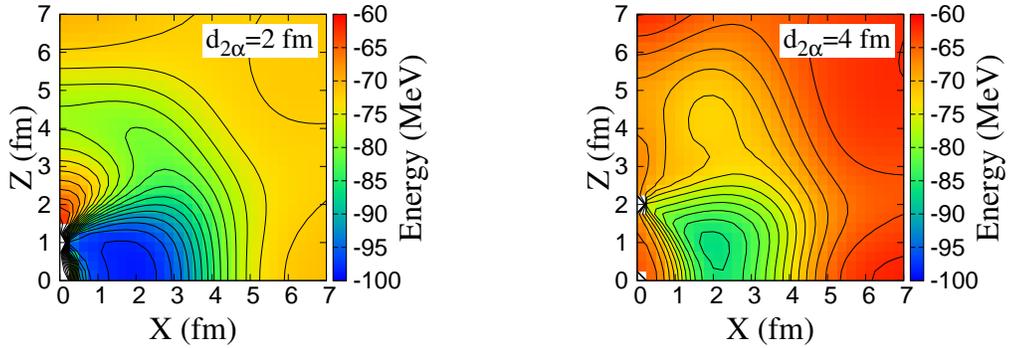}
  \end{center}
\caption{\label{fig:14C_ES}
$0^+$-projected energy surfaces for the  $^{10}$Be$(K=0)$+$\alpha$ wave function $\Phi^{0^+}_{^{10}{\rm Be}(K=0)+\alpha}(d_{2\alpha},D_{\alpha},\theta_{\alpha})$ plotted on the $X$-$Z$ plane.
The $\alpha$-$\alpha$ distances are chosen to be $d_{2\alpha}=2$ fm and $4$ fm.
}
\end{figure}

We discuss the energy of the $^{10}$Be+$\alpha$ wave function $\Phi_{^{10}{\rm Be}+\alpha}(d_{2\alpha},D_{\alpha},\theta_{\alpha},\phi)$ with respect to the $\alpha_3$ position around the $^{10}$Be core.
The $0^+$-projected energy is calculated for the $K=0$ projected $^{10}$Be core ($^{10}{\rm Be}(K=0)$+$\alpha$ wave function) as 
\begin{eqnarray}
\Phi^{0^+}_{^{10}{\rm Be}(K=0)+\alpha}(d_{2\alpha},D_{\alpha},\theta_{\alpha})
 &=&{\hat{P}^{0^+}_{00}} \sum_\phi \Phi_{^{10}{\rm Be}+\alpha}(d_{2\alpha},D_{\alpha},\theta_{\alpha},\phi), \label{eq:10Be(K=0)+a}
\end{eqnarray}
where the angle $\phi$ for the $nn$ position at eight points $\phi=\frac{\pi}{8}, \frac{3\pi}{8}, \frac{5\pi}{8}, \cdots,\frac{15\pi}{8}$ are summed with the equal weight to project the $^{10}$Be core to the $K=0$ eigen state.
Figure \ref{fig:14C_ES} shows the $0^+$ energy plotted on the $X$-$Z$ plane for the $\alpha_3$ position around the $^{10}{\rm Be}(K=0)$ core with $d_{2\alpha}=2$ fm and 4 fm.
In the energy surface for $d_{2\alpha}=2$ fm, we find the energy minimum around $(X,Z)=(2, 0)$ fm which corresponds to the ground state configuration of $^{14}$C.
Along the $Z$ axis, the energy is relatively high in the $D_{\alpha}<3.5$ fm region because of the Pauli blocking effect from another $\alpha$ cluster similarly to $^{12}$C.
In the $D_{\alpha}\sim 4$ fm region, an energy valley is seen toward $\theta_{\alpha}=0^{\circ}$, which corresponds to the linearly aligned $3\alpha$ on the $Z$ axis.
Comparing the energy surfaces for $^{14}$C with those for $^{12}$C, we find significant differences. 
The energy pocket for the ground state is deeper and the energy valley exists in the smaller $D_{\alpha}$ region in $^{14}$C than in $^{12}$C.
It means that valence neutrons effectively give an additional attraction between $\alpha$ clusters and keep three clusters in a compact region compared with the $3\alpha$ system without valence neutrons.

\begin{figure}[htbp]
\begin{center}
\includegraphics[width=12.0cm]{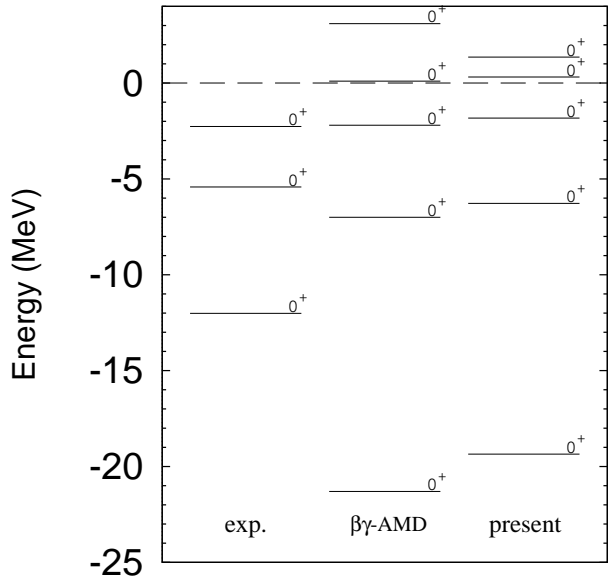}
\end{center}
\caption{\label{fig:level_14C}
$0^+$ energy spectra of $^{14}$C obtained by the Full-GCM calculation of the $^{10}$Be+$\alpha$-cluster model measured from the $^{10}$Be+$\alpha$ threshold energy.
The experimental data \cite{Exp-14C} and the theoretical spectra of the $\beta\gamma$-AMD \cite{Suhara-14C} are also shown.
}
\end{figure}

\subsection{GCM calculation of $^{14}$C}

We perform the $^{10}{\rm Be}+\alpha$ GCM calculation and obtain the ground and excited $0^+$ states of $^{14}$C by superposing the $^{10}$Be+$\alpha$ wave functions.
Figure \ref{fig:level_14C} shows the calculated $0^+$ energy spectra of $^{14}{\rm C}$ compared with the experimental data as well as the theoretical results of Ref.~\cite{Suhara-14C} calculated using the $\beta\gamma$-AMD by Suhara {\it et al.}
In the present calculation, we obtained the $0^+_4$ state having the dominant linear-chain structure slightly above the $^{10}$Be+$\alpha$ threshold energy.
It corresponds to the linear-chain state predicted as the $0^+_5$ state with the $\beta\gamma$-AMD.
Below this state, we obtain only two excited $0^+$ states. 
Compared with the $\beta\gamma$-AMD calculation, an excited $0^+$ state below the linear-chain state is missing in the present calculation, maybe, because $\alpha$ cluster breaking is omitted in the present model space of the $^{10}$Be+$\alpha$ wave functions.

Experimentally, only three states below the $^{10}$Be+$\alpha$ threshold energy have been confirmed to be $0^+$ states.
The ground state is overbound from the $^{10}$Be+$\alpha$ threshold in the present calculation similarly to the $\beta\gamma$-AMD calculation. 
This overbinding problem might come from the effective interaction of the two-body density-independent central force used in the two calculations.

\begin{figure}[htbp]
  \begin{center}
   \includegraphics[width=\hsize]{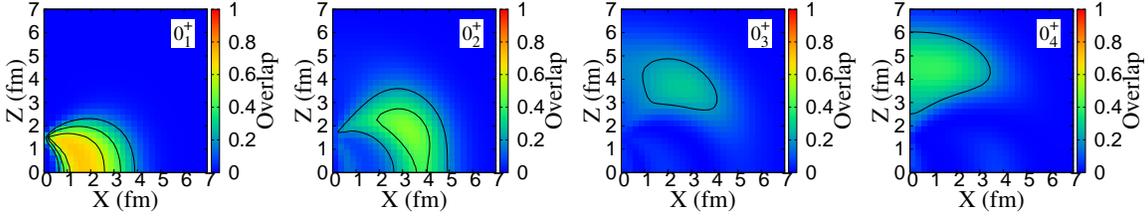}
  \end{center}
\caption{\label{fig:14C_Ov}
Overlaps with the $^{10}$Be+$\alpha$ wave functions ($d_{2\alpha}=3$ fm) for $^{14}$C$(0^+_1)$, $^{14}$C$(0^+_2)$, $^{14}$C$(0^+_3)$, and $^{14}$C$(0^+_4)$ obtained by the Full-GCM calculation of the $^{10}$Be+$\alpha$ model.
}
\end{figure}

To measure the $\alpha_3$ probability around the $^{10}{\rm Be}$, we calculate the overlap between the obtained GCM wave function for $^{14}$C($0^+_n$) and the $K=0$ projected $^{10}{\rm Be}$+$\alpha$ cluster wave function as 
\begin{eqnarray}
O^{0^+_n}_{^{10}{\rm Be}(K=0)+\alpha}(d_{2\alpha},D_{\alpha},\theta_{\alpha}) &=& 
\left| \braket{\Phi^{0^+}_{^{10}{\rm Be}(K=0)+\alpha} (d_{2\alpha},D_{\alpha},\theta_{\alpha})| \Psi^{0^+_n}_{^{14}{\rm C}}} \right|^2.
\label{eq:overlap_14C}
\end{eqnarray}
Figure \ref{fig:14C_Ov} shows the $\alpha_3$ probability in $^{14}{\rm C}(0^+_n)$ plotted on the $X$-$Z$ plane for the $\alpha_3$ position around the $^{10}{\rm Be}(K=0)$ with $d_{2\alpha}=3$ fm.
The $^{14}$C ground state shows the compact structure with the large $\alpha_3$ probability around $(X,Z)=(1.5, 0.75)$ fm.
In $^{14}$C($0^+_2$), the $\alpha_3$ probability is distributed in the larger $D_\alpha$ region than $^{14}$C($0^+_1$), but the spatial development of clustering is not as remarkable as $^{12}$C($0^+_2$). 
$^{14}$C($0^+_3$) has only small overlap with the $^{10}{\rm Be}(K=0)$+$\alpha$ wave function meaning that major component of this state is not the $^{10}{\rm Be}(K=0)$+$\alpha$ configuration.
$^{14}$C($0^+_4$) has the large $\alpha_3$ probability near the $Z$-axis in $3 {\rm \ fm}<Z<5 {\rm \ fm}$ region.
The largest probability exists around $(X,Z)=(0, 4.5)$ fm corresponding to the linear-chain $3\alpha$ structure.
In comparison with $^{12}$C, the probability in $^{14}$C($0^+_4$) is more concentrated on the $Z$-axis, i.e., the fluctuation against the bending motion is weaker than $^{12}$C($0^+_3$). 

In the present GCM calculation, we take into account the $\alpha$-cluster motion around the $^{10}$Be core, which is not sufficiently considered in the work by Suhara {\it et al.} \cite{Suhara-14C}.
Nevertheless, we obtain the linear-chain $3\alpha$ structure having the cluster structure quite similar to that predicted in Ref. \cite{Suhara-14C}.
The linear-chain structure in $^{14}$C is stable against the bending mode as well as the escaping mode of the $\alpha$ cluster owing to the existence of valence neutrons.
On the other, we find no excited state showing the gas-like feature of $3\alpha$ in $^{14}$C.
The reason is that $\alpha$ clusters are more strongly attracted by valence neutrons in $^{14}$C than $^{12}$C.
Below the linear-chain state, we obtain $^{14}{\rm C}(0^+_2$) and $^{14}{\rm C}(0^+_3$) with characters different from $^{12}{\rm C}(0^+_2)$.
More detailed discussions are given in the next section.

\begin{table}[htb]
\caption{  \label{table:14C}
Monopole transition strengths and rms radii for $0^+$ states of $^{14}$C.
Experimental data are from Ref. \cite{Exp-14C}.
}
  \centering
  \begin{tabular*}{11cm}{@{\extracolsep{\fill}} c|ccc}
  \hline
   & \multicolumn{2}{c}{Cal.} & \multicolumn{1}{c}{Exp.} \\
\cline{2-3} \cline{4-4}
   State & $B$(IS0) (fm$^4$) & $R_{\rm rms}$ (fm) &  $R_{\rm rms}$ (fm)\\ 
\hline
   $0^+_1$  &      & 2.38 &  2.30$\pm$0.07 \\ 
   $0^+_2$  & 34.7 & 2.67 &  \\ 
   $0^+_3$  & 54.8 & 2.81 &  \\ 
   $0^+_4$  & 95.2 & 2.90 &  \\
  \hline
  \end{tabular*}
\end{table}

In Table \ref{table:14C}, we show the monopole transition strengths and rms radii of $^{14}{\rm C}(0^+_n)$.
We do not find remarkable monopole transitions for excited $0^+$ states in $^{14}$C as that of $^{12}{\rm C}(0^+_2)$.
Among $^{14}$C($0^+_{2,3,4}$), $^{14}$C($0^+_4$) has significant monopole transition strength comparable to that of $^{12}$C($0^+_3$).
The calculated rms radius of the ground state reasonably reproduces the experimental radius.
The rms radii of the excited $0^+$ states $^{14}{\rm C}(0^+_{2,3,4})$ are longer than $^{14}$C($0^+_1$) but they are smaller than those of $^{12}{\rm C}(0^+_{2,3})$.
This indicates again that valence neutrons attract $\alpha$ clusters and suppress the spatially development of the 3$\alpha$ clustering.


\section{DISCUSSION}
\label{sec:discussion}

In this section, we discuss details of cluster modes in excited $0^+$ states of $^{12}$C and $^{14}$C such as the $\alpha_3$-cluster rotation $(\theta_\alpha)$ mode around Be cores and the $\alpha$-$\alpha$ ($d_{2\alpha}$) mode in Be cores and their effects on the monopole transition strengths by performing truncated GCM calculations within restricted model spaces.

\subsection{Effects of $\alpha_3$-cluster rotation around Be cores: analysis of fixed-$\theta$ GCM}

\begin{figure}[htbp]
\begin{center}
\includegraphics[width=\hsize]{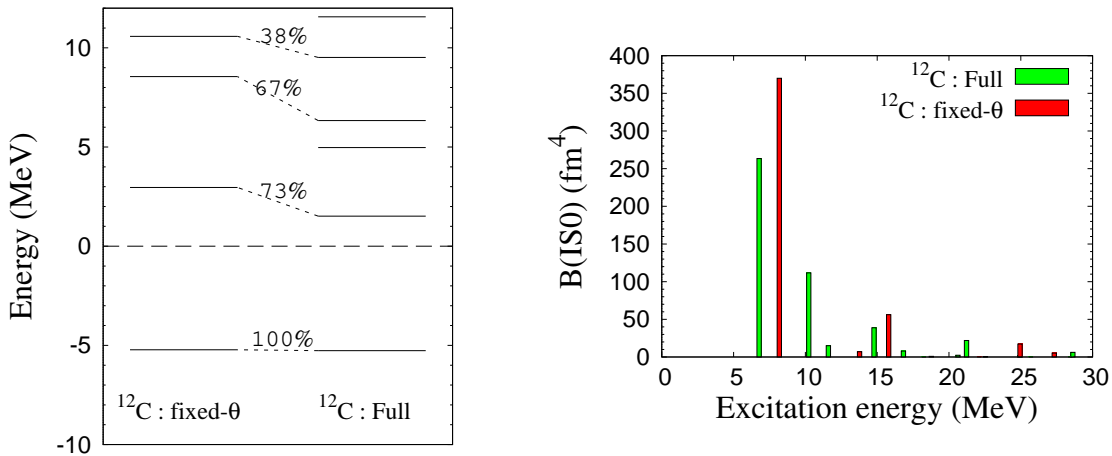}
\end{center}
\caption{\label{fig:12C_mode_theta}
$0^+$ energy spectra and monopole transition strengths of $^{12}$C obtained by the fixed-$\theta$ and Full-GCM calculations.
The energy levels with significant ($\gtrsim30\%$) overlaps between two calculations are connected by dotted lines. 
}
\end{figure}

To discuss effects of the $\alpha_3$-cluster rotation around the $^{8}$Be core in $^{12}$C, we preform the GCM calculation without the $\alpha_3$-cluster rotation by fixing $\theta_\alpha=\pi/2$ of the $3\alpha$ wave functions (named fixed-$\theta$ GCM).
Namely, we use only two parameters, $d_{2\alpha}$ and $D_\alpha$, as the generator coordinates in the fixed-$\theta$ GCM.
Note that the angular motion of the $\alpha_3$ cluster is equivalent to the $^{8}$Be core rotation in a $J^\pi$ eigen state.
We compare the fixed-$\theta$ GCM (without rotation) with the Full-GCM (with rotation).
In Fig. \ref{fig:12C_mode_theta}, we show the $0^+$ energy spectra and monopole transition strengths of $^{12}{\rm C}$ calculated by the fixed-$\theta$ GCM compared with the Full-GCM. 
We also show squared overlaps of the wave functions for the $0^+$ states between two calculations.
It is found that the $0^+_1$ state of the fixed-$\theta$ GCM (fixed-$\theta$ $0^+_1$) has 100\% overlap with that of the Full-GCM (Full-GCM $0^+_1$) indicating that the $\alpha_3$ rotation gives almost no contribution to the ground state.
On the other hand, the rotation effect gives important contributions to excited $0^+$ states.
The fixed-$\theta$ $0^+_2$ has 73\% overlap with the Full-GCM $0^+_2$ and gains significant energy by the rotation to form the gas-like state.
For the Full-GCM $0^+_3$, there is no corresponding state in the fixed-$\theta$ GCM because the bending chain $3\alpha$ structure is mainly described by $\theta \sim 0 $ configurations, which are omitted in the fixed-$\theta$ GCM.
It should be noted that the fixed-$\theta$ $0^+_2$ has 23\% overlap with the Full-GCM $0^+_3$. In other words, the component of the fixed-$\theta$ $0^+_2$ is split by the rotation into the Full-GCM $0^+_2$ and $0^+_3$ states with fractions of 73\% and 23\%, respectively.

Let us compare the monopole transition strengths of $^{12}$C between fixed-$\theta$ GCM and Full-GCM calculations.
In the fixed-$\theta$ GCM, the monopole transition strength is concentrated at the $0^+_2$ state.
The strong monopole transition for the fixed-$\theta$ $0^+_2$ is split by the rotation into the Full-GCM $0^+_2$ and $0^+_3$, consistently with the split of the fixed-$\theta$ $0^+_2$ component to these two states with the fractions of 73\% and 23\%. 
As a result, in the Full-GCM result, $^{12}$C($0^+_2$) has the remarkable monopole transition strength, whereas $^{12}$C($0^+_3$) has the relatively smaller but significant monopole transition strength.
The present result is consistent with the rotation effect in the monopole strength discussed by one of the authors in Ref.~\cite{Enyo-12C}.

\begin{figure}[htbp]
\begin{center}
\includegraphics[width=\hsize]{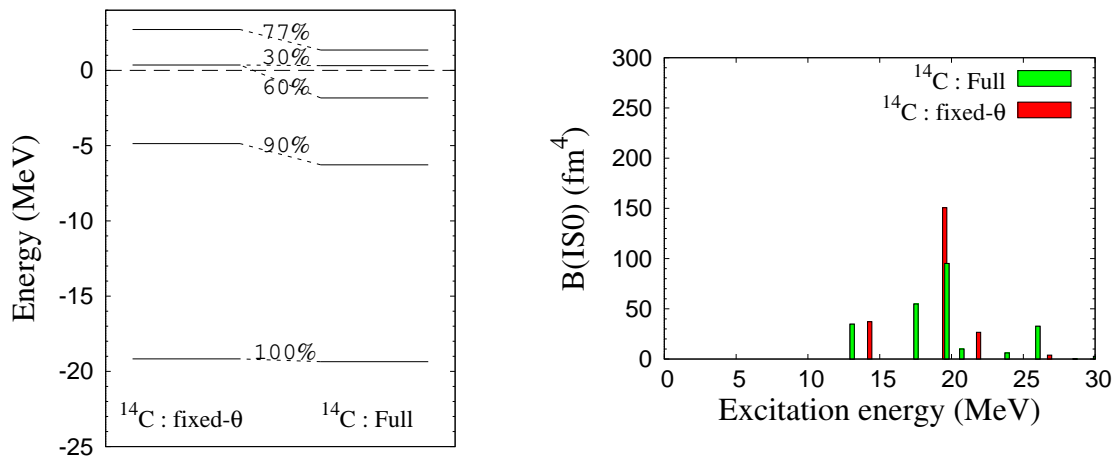}
\end{center}
\caption{\label{fig:14C_mode_theta}
$0^+$ energy spectra and monopole transition strengths of $^{14}$C obtained by the fixed-$\theta$ and Full-GCM calculations.
The energy levels with significant ($\gtrsim30\%$) overlap between two calculations are connected by dotted lines. 
}
\end{figure}

In a similar way to the analysis for $^{12}$C, we calculate the fixed-$\theta$ GCM for $^{14}$C to discuss the rotation effect of the $\alpha_3$ cluster around the $^{10}$Be core.
In the fixed-$\theta$ GCM for $^{14}$C, we use three parameters, $d_{2\alpha}$, $D_\alpha$, and $\phi$, as the generator coordinates and fix $\theta_\alpha=\pi/2$.
In Fig. \ref{fig:14C_mode_theta}, we show the $0^+$ energy spectra and monopole transition strengths of $^{14}{\rm C}$ calculated by the fixed-$\theta$ GCM compared with the Full-GCM, and also show the squared overlaps of the obtained wave functions between two calculations.
For $^{14}$C($0^+_1$), the $\alpha_3$ rotation gives almost no contribution (100\% overlap). 
For $^{14}$C($0^+_2$), the fixed-$\theta$ $0^+_2$ has 90\% overlap with the Full-GCM $0^+_2$ and gains about $1$ MeV energy by the rotation.
For $^{14}$C($0^+_3$) of the Full-GCM, the fixed-$\theta$ $0^+_3$ has the dominant overlap as 60\%.
For the linear-chain state, $^{14}$C($0^+_4$) obtained by the Full-GCM, there is no corresponding state in the fixed-$\theta$ GCM because it contains dominantly the $\theta \sim 0 $ components, which are omitted in the fixed-$\theta$ GCM. 
It should be noted that the fixed-$\theta$ $0^+_3$ component is fragmented by the rotation into the Full-GCM $0^+_3$ and $0^+_4$ states with fractions of 60\% and 30\%, respectively.

In the monopole transition strengths calculated by the fixed-$\theta$ GCM, the strength is concentrated at the $0^+_3$ state at $E_x\sim 20$ MeV, but the absolute value of $B$(IS0) for this state is about half of the fixed-$\theta$ $0^+_2$ state of $^{12}$C.
As a result of rotation, the monopole strength concentrated at the fixed-$\theta$ $0^+_3$ is fragmented into the Full-GCM $0^+_3$ and $0^+_4$ in this energy region.
However the ratio of the monopole strengths for the Full-GCM $0^+_3$ and $0^+_4$ is not consistent with the fractions (60\% and 30\%) of the fixed-$\theta$ $0^+_3$ component in these two states.
Namely, even though the Full-GCM $0^+_3$ contains the significant fixed-$\theta$ $0^+_3$ component, it has the relatively weaker monopole transition than the Full-GCM $0^+_4$.  
In order to understand cluster structures and their relation to the monopole strengths in $^{14}$C, further detailed analysis is necessary by taking into account valence neutron configurations.

\subsection{Valence neutron configurations in $^{14}$C}

In this section, we discuss the effect of valence neutron mode ($\phi$) in the cluster structures of $^{14}$C.
As described previously, we adopt 8 points $\phi=\pi/8, 3\pi/8, 5\pi/8, 7\pi/8, 9\pi/8, 11\pi/8, 13\pi/8, $ 
$15\pi/8$ for the $nn$ angle position around the $2\alpha$ in the $^{10}$Be core.
To see valence neutrons distribution in the $^{10}$Be core, we classify the neutron configurations into two groups. 
One is the set $\phi=\pi/8, 7\pi/8, 9\pi/8, 15\pi/8$ called vertical, in which two neutrons are located in a direction vertical to the $X$-$Z$ plane, and the other is the set $\phi_j=3\pi/8, 5\pi/8, 11\pi/8, 13\pi/8$ called planar, in which the $3\alpha$ and two neutrons form approximately planar configurations on the $X$-$Z$ plane.
Here, the $^{10}{\rm Be}({\rm Vertical})+\alpha$ and $^{10}{\rm Be}({\rm Planar})+\alpha$ wave functions of vertical and planar configurations projected onto $0^+$ states are defined as
\begin{eqnarray}
\Phi^{0^+}_{^{10}{\rm Be}({\rm vertical})+\alpha}(d_{2\alpha}, D_{\alpha}, \theta_{\alpha})&=&
\sum_{\phi \in {\rm vertical}} {\hat{P}^{0^+}_{00}} \Phi_{^{10}{\rm Be}+\alpha}(d_{2\alpha}, D_{\alpha}, \theta_{\alpha}, \phi),
\\
\Phi^{0^+}_{^{10}{\rm Be}({\rm planar})+\alpha}(d_{2\alpha}, D_{\alpha}, \theta_{\alpha})&=&
\sum_{\phi \in {\rm planar}} {\hat{P}^{0^+}_{00}} \Phi_{^{10}{\rm Be}+\alpha}(d_{2\alpha}, D_{\alpha}, \theta_{\alpha}, \phi).
\end{eqnarray}

\begin{figure}[htbp]
  \begin{center}
   \includegraphics[width=\hsize]{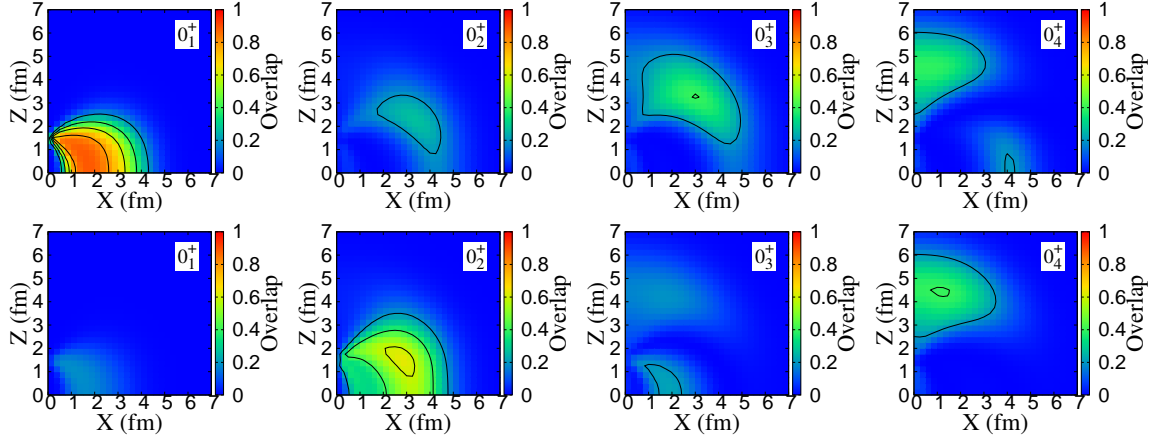}
  \end{center}
\caption{\label{fig:14C_Ov_vp}
(Upper) overlaps with the $^{10}{\rm Be}({\rm vertical})+\alpha$ wave function ($d_{2\alpha}=3$ fm) for $^{14}$C($0^+_{1,2,3,4})$ obtained by the Full-GCM.
(Lower) overlaps with the $^{10}{\rm Be}({\rm planar})+\alpha$ wave function ($d_{2\alpha}=3$ fm) for $^{14}$C($0^+_{1,2,3,4})$ obtained by the Full-GCM.
}
\end{figure}

We calculate the $^{10}{\rm Be}({\rm vertical})+\alpha$ and $^{10}{\rm Be}({\rm planar})+\alpha$ components in the $0^+$ states of $^{14}$C obtained by the Full-GCM wave function.
In Fig.\ref{fig:14C_Ov_vp}, the vertical and planar components in $^{14}{\rm C}(0^+_n)$ are shown on the $X$-$Z$ plane for the $\alpha_3$ position around the $^{10}$Be core with $d_{2\alpha}=3$ fm.
We find the vertical component is dominant in the ground state because the planar configuration is suppressed in such the compact cluster state because of the Pauli blocking from the $\alpha_3$ cluster for valence neutrons in the planar configuration.
On the other hand, the planar configuration is dominant in $^{14}$C($0^+_2$).
This state has a somewhat enhanced cluster structure, in which the valence neutrons in the planar configurations gain much potential energy compared with those in the vertical configurations.
This result is consistent with the discussion in Ref.~\cite{Suhara-14C}.
In $^{14}$C($0^+_3$), the vertical component is dominant whereas the planar one is minor.
We find the linear-chain $3\alpha$ state at $^{14}$C($0^+_4$) has almost the same amount of the vertical and planar components in the region near the $Z$-axis.
However, $^{14}$C($0^+_4$) also has the significant (about 20\%) vertical component around $(X,Z)=(4,0)$ fm.
As shown later, this vertical component contributes to the monopole strength of $^{14}$C($0^+_4$).

\begin{figure}
\begin{center}
\includegraphics[width=\hsize]{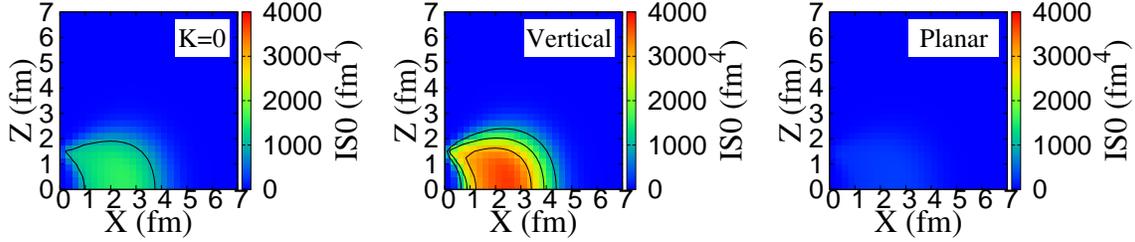}
\end{center}
\caption{\label{fig:14C_mono_ov}
Monopole transition strengths from the approximated ground state wave function of $^{14}$C to the $^{10}{\rm Be}(K=0)+\alpha$ wave functions, those to the $^{10}{\rm Be}({\rm Vertical})+\alpha$ wave functions, and those to the $^{10}{\rm Be}({\rm Planar})+\alpha$ wave functions. 
}
\end{figure}

Let us discuss which configurations of $\Phi^{0^+}_{^{10}{\rm Be}(K=0)+\alpha}$, $\Phi^{0^+}_{^{10}{\rm Be}({\rm vertical})+\alpha}$ and $\Phi^{0^+}_{^{10}{\rm Be}({\rm planar})+\alpha}$ are directly excited by the monopole operator from the ground state.
We calculate the monopole transition from $^{14}$C$(0^+_1)$ to a single $^{10}$Be+$\alpha$ configuration.
For simplicity, we approximate the ground state wave function with the $^{10}$Be+$\alpha$ wave function with fixed parameters $d_{2\alpha}=3$ fm, $D_{\alpha}=2$ fm, and $\theta_{\alpha}=\pi/2$, by superposing $\phi=\frac{\pi}{8},\frac{3\pi}{8},\frac{5\pi}{8},\cdots,\frac{15\pi}{8}$ bases.
The approximated ground state wave function $\Psi^{0^+_1}_{^{14}{\rm C}, {\rm app}}$ has 93\% overlap with the ground state of the Full-GCM.
The matrix element for the monopole transition from $^{14}$C$(0^+_1)$ to the vertical, planar, and $^{10}{\rm Be}(K=0)+\alpha$ configurations are given as 
\begin{eqnarray}
B({\rm IS0})_{^{10}{\rm Be}({\rm Vertical})+\alpha}(d_{2\alpha},D_{\alpha},\theta_{\alpha}) &=&
\left| \braket{\Psi^{0^+_1}_{^{14}{\rm C}, {\rm app}}|\mathcal{M}({\rm IS0})|\Phi^{0^+}_{^{10}{\rm Be}({\rm vertical})+\alpha} (d_{2\alpha},D_{\alpha},\theta_{\alpha})} \right|^2,  \nonumber \\
\label{eq:mono_overlap_v14C} \\
B({\rm IS0})_{^{10}{\rm Be}({\rm Planar})+\alpha}(d_{2\alpha},D_{\alpha},\theta_{\alpha}) &=&
\left| \braket{\Psi^{0^+_1}_{^{14}{\rm C}, {\rm app}}|\mathcal{M}({\rm IS0})|\Phi^{0^+}_{^{10}{\rm Be}({\rm planar})+\alpha} (d_{2\alpha},D_{\alpha},\theta_{\alpha})} \right|^2,  \nonumber \\
\label{eq:mono_overlap_p14C} \\
B({\rm IS0})_{^{10}{\rm Be}(K=0)+\alpha}(d_{2\alpha},D_{\alpha},\theta_{\alpha}) &=& 
\left| \braket{\Psi^{0^+_1}_{^{14}{\rm C}, {\rm app}}|\mathcal{M}({\rm IS0})|\Phi^{0^+}_{^{10}{\rm Be}(K=0)+\alpha} (d_{2\alpha},D_{\alpha},\theta_{\alpha})} \right|^2.  \nonumber \\
\label{eq:mono_overlap_k14C}
\end{eqnarray}
In Fig.~\ref{fig:14C_mono_ov}, the monopole strengths in $^{14}$C for the three kinds of configurations are plotted on the $X$-$Z$ plane. 
We find that the vertical configurations in the $\theta_{\alpha} \sim \pi/2$ region are directly excited by the monopole operator from the ground state, because the ground state dominantly consists of compact vertical configurations with $\theta_\alpha \sim \pi/2$ as shown in Fig. \ref{fig:14C_Ov_vp}.
Since the monopole operator does not cause rotation, it cannot excite planar configurations. 
It means that $^{14}{\rm C}(0^+_n)$ are excited by the monopole operator mainly through the vertical component in the $\theta_{\alpha} \sim \pi/2$ region.

Let us consider an additional truncation of the model space in the GCM calculation to discuss effects of valence neutron configurations in the $^{10}$Be core on the cluster structures and monopole strengths in $^{14}$C.
We perform the GCM calculation using only the vertical basis wave functions (V-GCM), in which we superpose the vertical wave functions $\Phi^{0^+}_{^{10}{\rm Be}({\rm Vertical})+\alpha}(d_{2\alpha}, D_{\alpha}, \theta_{\alpha})$.
For the generator coordinates, $d_{2\alpha}$, $D_{\alpha}$ and $\theta_{\alpha}$, we use the same parametrization as the Full-GCM; $d_{2\alpha}=2, 3, 4$ fm, $D_{\alpha}=2, 3, \cdots, 7$ fm and $\theta_{\alpha}=0, \pi/8, \pi/4, 3\pi/8, \pi/2$.

\begin{figure}[htbp]
\begin{center}
\includegraphics[width=\hsize]{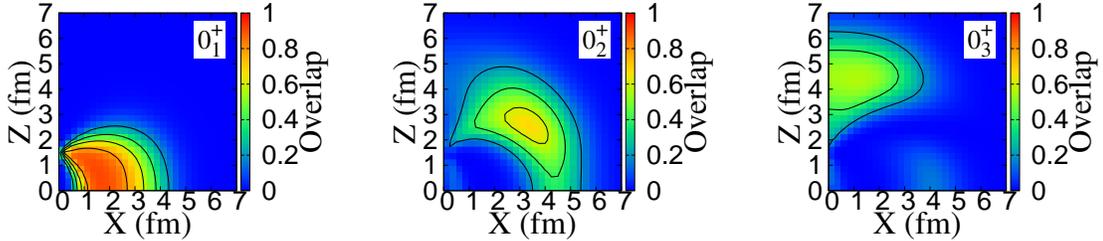}
\end{center}
\caption{\label{fig:14C_Ov_vv}
Overlaps with the $^{10}$Be(vertical)+$\alpha$ wave function for $0^+_1$, $0^+_2$, $0^+_3$, and $0^+_4$ obtained by the V-GCM calculation.  
The overlaps are plotted on the X-Z plane for the $\alpha_3$ position around the $^{10}$Be(vertical) core with $d_{2\alpha}=$ 3 fm.
}
\end{figure}

In order to discuss $3\alpha$ dynamics in the model space of the vertical configuration, we show, in Fig.~\ref{fig:14C_Ov_vv}, the overlaps with the $^{10}$Be(Vertical)+$\alpha$ wave function on the $X-Z$ plane for the $0^+$ states obtained by the V-GCM calculation.
The $0^+_1$ state has a compact structure.
In the $0^+_2$ state, the $\alpha_3$ probability is spread widely in the larger $D_\alpha$ region than the $0^+_1$ state and has a gas-like feature, although the spatial development of clustering is not as prominent as $^{12}{\rm C}(0^+_2)$ because of the attraction from the valence neutrons in the vertical configuration.
The $0^+_3$ state shows a bending-chain $3\alpha$ structure similar to $^{12}{\rm C}(0^+_3)$, but the cluster development is weaker than that of $^{12}{\rm C}(0^+_3)$.

\begin{figure}[htbp]
\begin{center}
\includegraphics[width=\hsize]{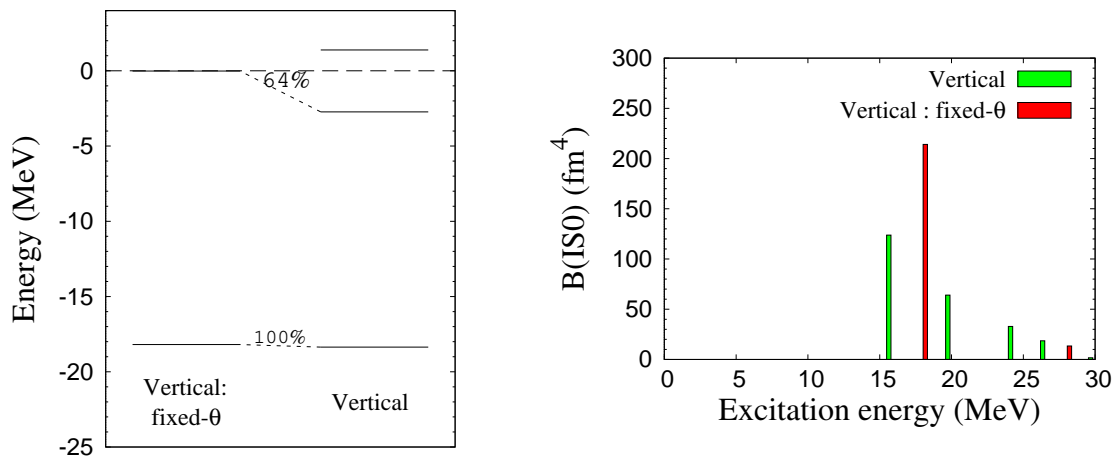}
\end{center}
\caption{\label{fig:v14C_mode}
$0^+$ energy spectra and monopole transition strengths of $^{14}$C obtained by the fixed-$\theta$ V-GCM and V-GCM calculations.
The energy levels with significant ($\gtrsim30\%$) overlap between two calculations are connected by dotted lines.
}
\end{figure}

To discuss effects of the $\alpha_3$-cluster rotation around the $^{10}$Be(vertical) core, we preform the fixed-$\theta$ GCM calculation in the vertical configurations, called fixed-$\theta$ V-GCM and compare the result with the V-GCM result.
Figure \ref{fig:v14C_mode} shows the $0^+$ energy spectra and monopole transition strengths of $^{14}$C calculated by the fixed-$\theta$ V-GCM and the V-GCM as well as the overlap of the obtained wave functions.

The $0^+_1$ state of the fixed-$\theta$ V-GCM has 100\% overlap with that of the V-GCM.
The $0^+_2$ state obtains about 2 MeV energy gain by the $\alpha_3$ rotation.
For the $0^+_3$ state of the V-GCM, there is no corresponding state in the fixed-$\theta$ V-GCM, because it has a bending-chain $3\alpha$ structure like that of $^{12}{\rm C}(0^+_3)$ and contains dominantly $\theta_\alpha\sim 0$ configurations, which are omitted in the fixed-$\theta$ V-GCM.
It should be commented that the component of the fixed-$\theta$ V-GCM $0^+_2$ is fragmented by the rotation into the V-GCM $0^+_2$ and $0^+_3$ states with fractions of 64\% and 25\% .

In the monopole strengths, the strongly concentrated strength can be seen at the $0^+_2$ state in the fixed-$\theta$ V-GCM, and it is split by the rotation to the $0^+_2$ and $0^+_3$ states in the V-GCM.
The ratio of the monopole strengths of the V-GCM $0^+_2$ and $0^+_3$ states is consistent with the ratio of the component of the fixed-$\theta$ V-GCM $0^+_2$ state in these two states. 
Namely, as a result of the dominant fixed-$\theta$ V-GCM $0^+_2$ component in the V-GCM $0^+_2$, the V-GCM $0^+_2$ has the largest monopole strength.

These results indicate that features of the cluster structures and monopole strengths in $^{14}$C in the vertical model space are quite similar to those of $^{12}$C, though some quantitative differences are seen in the weaker cluster development and monopole transitions in $^{14}$C of the V-GCM than $^{12}$C.
It means that 3$\alpha$ clusters with the valence neutrons in the vertical configuration form cluster structures similar to the 3$\alpha$ system without valence neutrons except for the somewhat weaker spatial extend of cluster motion because of the additional attraction by the valence neutrons in the vertical configuration. 
Therefore, the qualitative differences of cluster features of the Full-GCM between $^{14}$C and $^{12}$C originate in coupling of the vertical with planar configurations.

\begin{figure}[htbp]
\begin{center}
\includegraphics[width=\hsize]{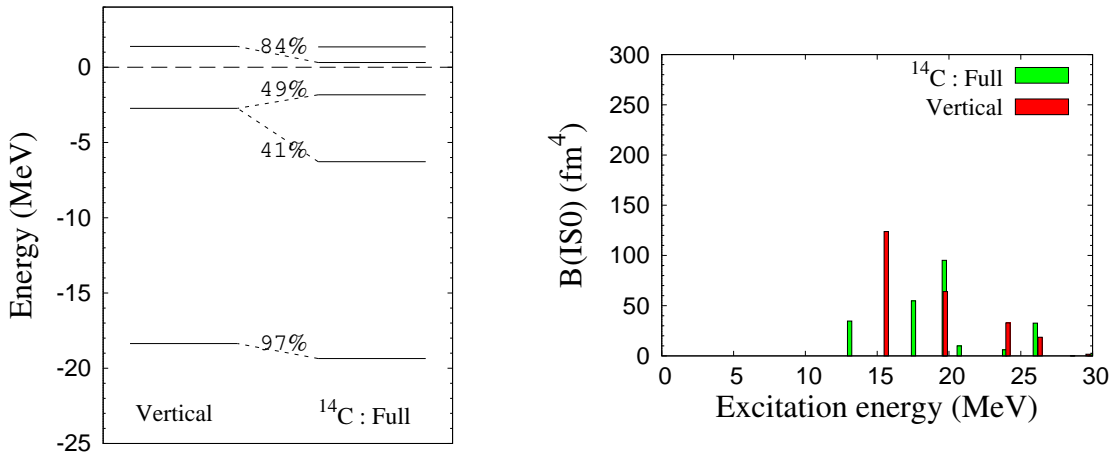}
\end{center}
\caption{\label{fig:vf14C_mode}
$0^+$ energy spectra of $^{14}$C and monopole transition strengths obtained by the V-GCM and Full-GCM calculations.
The energy levels with significant ($\gtrsim30\%$) overlap between two calculations are connected by dotted lines.
}
\end{figure}

We compare the V-GCM result with the Full-GCM one to discuss the coupling effect of the vertical and planar configurations.
Figure \ref{fig:vf14C_mode} shows the $0^+$ energy spectra and monopole transition strengths of $^{14}$C of the V-GCM and Full-GCM as well as the overlap of the obtained wave functions.
The coupling gives almost no contribution to the ground state.
The V-GCM $0^+_2$ state is split into two states ($0^+_2$ and $0^+_3$) in the Full-GCM because of the coupling with the planar configurations.
The linear-chain state at $^{14}$C($0^+_4$) of the Full-GCM has the major overlap of 84\% with the V-GCM $0^+_3$ state having the bending-chain $3\alpha$ structure.
It means that the bending-chain $3\alpha$ structure is stabilized to form the linear chain 3$\alpha$ structure because of the coupling of the vertical and planar configurations.

In the monopole transition strengths, the V-GCM $0^+_2$ has the strongest transition reflecting the large $^{10}{\rm Be}({\rm Vertical})+\alpha$ component in the $\theta_\alpha \sim \pi/2$ region, which can be directly excited by the monopole operator from the ground state as discussed previously.
However, the coupling with the planar configurations fragments, the monopole transition strengths of the V-GCM $0^+_2$ state into the Full-GCM $0^+_2$ and $0^+_3$ states, and also somewhat enhances the monopole transition strength of the Full-GCM $0^+_4$ state.
As a result, there is no concentration of the monopole strengths in this energy region. 
$^{14}$C($0^+_4$) of the Full-GCM has the monopole strength comparable to $^{12}$C($0^+_3$) because it contains the significant $^{10}{\rm Be}({\rm Vertical})+\alpha$ component in the $\theta_\alpha \sim \pi/2$ region, which is regarded as the remains of the bending-chain structure in the fixed-$\theta$ V-GCM $0^+_3$.

\subsection{Effects of $\alpha$-$\alpha$ mode of $^{8}$Be and $^{10}$Be cores}

\begin{figure}[htbp]
\begin{center}
\includegraphics[width=\hsize]{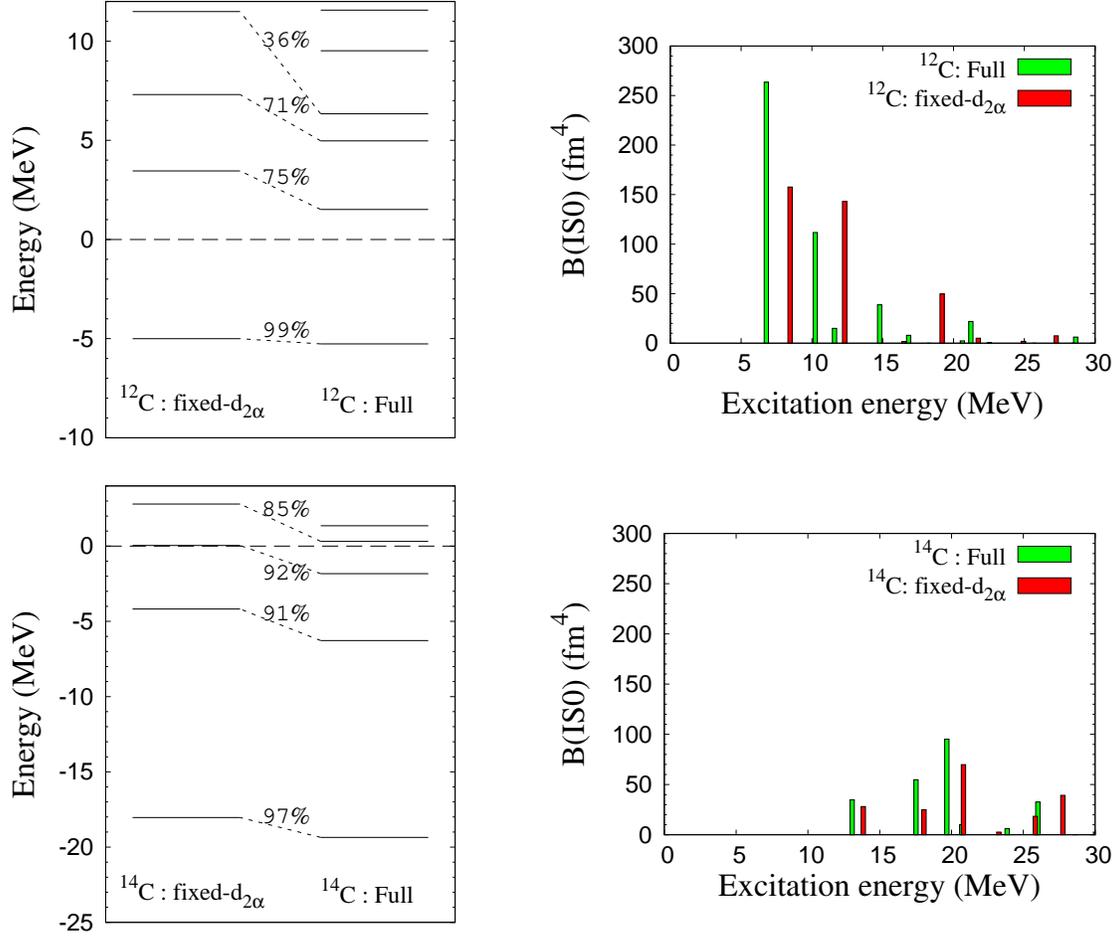}
\end{center}
\caption{\label{fig:level_C_mode}
$0^+$ energy spectra and monopole strengths of $^{12}$C and $^{14}$C calculated by the fixed-$d_{2\alpha}$ GCM compared with those of the Full-GCM.
The energy levels with significant ($\gtrsim30\%$) overlap between two calculations are connected by dotted lines.
}
\end{figure}

As discussed in Sec. \ref{sec:results-Be}, the $^{8}$Be core is soft, but the $^{10}$Be core is relatively stiff against the $\alpha$-$\alpha$ mode.
To see the effect of the $\alpha$-$\alpha$ mode of the Be cores in $^{12}$C and $^{14}$C, we perform the GCM calculations with a fixed $\alpha$-$\alpha$ distance, $d_{2\alpha}=3$ fm (named fixed-$d_{2\alpha}$ GCM).
In Fig.~\ref{fig:level_C_mode}, we show the $0^+$ energy spectra and monopole strengths in $^{12}$C and $^{14}$C calculated by the fixed-$d_{2\alpha}$ GCM compared with those of the Full-GCM. 
We find significant energy gains in $^{12}{\rm C}(0^+_2)$, $^{12}{\rm C}(0^+_3)$, and $^{12}{\rm C}(0^+_4)$ because of the $\alpha$-$\alpha$ mode in the $^{8}$Be core.
Moreover, the monopole strength for $^{12}{\rm C}(0^+_2)$ is remarkably enhanced by the soft $\alpha$-$\alpha$ mode.
In contrast to the significant effects of the $\alpha$-$\alpha$ mode in $^{12}$C, the effects of the $\alpha$-$\alpha$ mode in $^{14}$C is not so remarkable, the fixed-$d_{2\alpha}$ GCM results are qualitatively similar to the Full-GCM for $^{14}$C.
It is concluded that the $\alpha$-$\alpha$ mode plays an important role in the lowered energy and the enhanced monopole strength of $^{12}{\rm C}(0^+_2)$, whereas it is less important in $^{14}$C.

\subsection{Cluster features of $^{12}$C and $^{14}$C}
As shown previously, in $^{12}$C, the $0^+_2$ shows the cluster gas-like feature, and the $0^+_3$ has the bending-chain structure.
In $^{14}$C, the linear-chain structure is obtained in the $0^+_4$ state, but a cluster gas-like structure does not appear in low-lying $0^+$ states.
In the previous sections, we analyzed cluster modes such as the $\alpha_3$-cluster rotation mode ($\theta_\alpha$ mode) around Be cores, the $nn$ rotation mode ($\phi$ mode) around the $2\alpha$ in the $^{10}$Be core of $^{14}$C, and the $\alpha$-$\alpha$ mode ($d_{2\alpha}$ mode) in Be cores, and discussed their contributions on energy spectra and monopole transition strengths in $^{12}$C and $^{14}$C.
We here discuss cluster features in $^{12}$C and $^{14}$C based on the present analyses using the $3\alpha$- and $^{10}$Be+$\alpha$-cluster models, while focusing on the role of valence neutrons in $^{14}$C.

In general, valence neutrons in $^{14}$C attract $\alpha$ clusters and suppress the spatially development of the 3$\alpha$ clustering.
In comparison between the fixed-$\theta$ GCM and Full-GCM calculations, we found that the $\alpha_3$-cluster rotation (the $\theta_\alpha$ mode) is essential to form the bending chain $3\alpha$ structure in $^{12}$C and the linear chain 3$\alpha$ structure in $^{14}$C.
The $\theta_\alpha$ mode splits the fixed-$\theta$ GCM $0^+_2$ into the Full-GCM $0^+_2$ and $0^+_3$ in $^{12}$C.
Because of the $\theta_\alpha$ mode, the monopole strengths that are well concentrated at the fixed-$\theta$ GCM $0^+_2$ are also split into two states, the Full-GCM $0^+_2$ and $0^+_3$ in $^{12}$C, resulting in the remarkable monopole strength in $^{12}$C($0^+_2$) and relatively weak monopole strength in $^{12}$C($0^+_3$).
In $^{14}$C, the $\theta_\alpha$ mode splits the fixed-$\theta$ GCM $0^+_3$ into the Full-GCM $0^+_3$ and $0^+_4$. Differently from the case of $^{12}$C, the effect of the $\theta_\alpha$ mode on the monopole strengths cannot be understood simply because the $\theta_\alpha$ mode couples with the $nn$ rotation mode ($\phi$ mode).

In the analysis of the vertical and planar configurations in $^{14}$C, we find the cluster structures of the V-GCM $0^+_1$, $0^+_2$, and $0^+_3$ of $^{14}$C obtained within the vertical model space are qualitatively similar to those of $^{12}$C($0^+_1$), $^{12}$C($0^+_2$), and $^{12}$C($0^+_3$) obtained by the Full-GCM, although they are quantitatively different in the spatial development of $3\alpha$ clustering because valence neutrons attract the 3$\alpha$ clusters in $^{14}$C.
It indicates that the coupling of the vertical configuration with the planar one gives essential contribution to the qualitative differences in cluster structures between $^{14}$C and $^{12}$C.
The Full-GCM $0^+_2$ state in $^{14}$C originates mainly in the planar configuration.
Moreover, the bending-chain structure obtained in the V-GCM $0^+_3$ state in $^{14}$C is stabilized by the coupling with the planar configuration.
These roles of the planar configuration are consistent with the mechanism of the linear-chain structure in $^{14}$C discussed by Suhara {\it et al.}\cite{Suhara-14C}. 
It means that the $\phi$ mode, i.e., the $nn$ configuration in the $^{10}$Be core plays important roles in the cluster structures of $^{14}$C.
In $^{14}$C, because of the strong coupling with the $\phi$ mode in addition to that with the $\theta_\alpha$ mode, the monopole strengths are fragmented into many $0^+$ states and show no concentration on a single $0^+$ state.

In comparison between the fixed-$d_{2\alpha}$ GCM and Full-GCM calculations, we found that the $d_{2\alpha}$ mode significantly enhances the monopole strengths in $^{12}$C because the $^{8}$Be core is soft against the $d_{2\alpha}$ mode.
By contrast, it does not give a drastic enhancement of monopole strengths in $^{14}$C because valence neutrons in the $^{10}$Be core tightly bind the 2$\alpha$s and suppress the $\alpha$-$\alpha$ distance.

\section{SUMMARY}
\label{sec:Summary}

We studied cluster structures in $0^+$ states of $^{12}$C and $^{14}$C with the $3\alpha$- and the $^{10}$Be+$\alpha$-cluster models, respectively, and discussed monopole transitions.

In $^{12}$C, we obtained the gas-like $3\alpha$-cluster structure of the $0^+_2$ state and the bending-chain $3\alpha$ structure of the $0^+_3$ state, which are consistent with microscopic $3\alpha$-cluster model calculations.
In $^{14}$C, we found the linear-chain $3\alpha$ structure of the $0^+_4$ state near the $^{10}$Be+$\alpha$ threshold.
On the other hand, a cluster gas-like structure does not appear in $^{14}$C.
It was found that valence neutrons in $^{14}$C attract $\alpha$ clusters and suppress the spatial development of the $3\alpha$ clustering compared with the $3\alpha$-cluster structure in $^{12}$C.
The valence neutrons stabilize the linear-chain state in $^{14}$C against the bending mode and the escaping mode of the $\alpha$ cluster.

We also investigated monopole transitions in  $^{12}$C and $^{14}$C and analyzed effects of $\alpha$-cluster motion and $nn$ configurations on the monopole transitions. 
In $^{14}$C, because of the strong coupling of the $\alpha$-cluster motion and $nn$ configurations, monopole transition strengths are fragmented into many $0^+$ states. 
The monopole transition strengths in $^{12}$C are enhanced by the $\alpha$-$\alpha$ motion in the $^{8}$Be core, but however, those in $^{14}$C are not enhanced so much by the $\alpha$-$\alpha$ motion in the $^{10}$Be core because valence neutrons tightly bind 2$\alpha$ clusters in the core.


\section{Acknowledgements}

The numerical calculations in this work were carried out 
using supercomputers at the Yukawa Institute for theoretical physics, Kyoto University.
This work was supported by JSPS KAKENHI Grant Number 26400270.

\end{document}